\documentclass[paper]{JHEP3}

\usepackage{amsfonts}
\usepackage[centertags]{amsmath}
\usepackage{amssymb}
\usepackage{epsfig}
\usepackage{subfigure}
\usepackage[section]{placeins}
\providecommand{\mislash}[1]{#1 \mspace{-10.0mu} \slash}

\providecommand{\proarrow}[0]{\rightarrow}

\providecommand{\dif}[0]{\mathrm{d}}

\providecommand{\proname}[2]{#1 \proarrow #2}

\providecommand{\cst}[0]{\hat{\sigma}}
\providecommand{\abs}[1]{\left\lvert #1 \right\rvert}

\providecommand{\abss}[1]{\left\lvert #1 \right\rvert^2}

\providecommand{\miim}[1]{{\rm Im} \left[ #1 \right]}

\providecommand{\g}[2]{\gamma^{#1}_{#2}}

\providecommand{\gp}[2]{\gamma^{#1 \; '}_{#2}}

\providecommand{\gfe}[0]{\gamma_{\text{FCI}}}

\providecommand{\ih}[0]{ih}
\providecommand{\il}[0]{il}
\providecommand{\dosh}[0]{2h}
\providecommand{\dosl}[0]{2l}
\providecommand{\unoh}[0]{1h}
\providecommand{\unol}[0]{1l}
\providecommand{\mibr}[2]{\text{Br}(\proname{#1}{#2})}

\newcommand{\be}{\begin{equation}}
\newcommand{\ee}{\end{equation}}
\newcommand{\bea}{\begin{eqnarray}}
\newcommand{\eea}{\end{eqnarray}}

\title{Leptogenesis with small violation of $B-L$}

\author{J.~Racker\footnote{racker@ific.uv.es}\\
Depto.\ de F\'{\i}sica Te\'orica and
IFIC, Universidad de
Valencia-CSIC \\ 
Edificio de Institutos de Paterna, Apt. 22085, 46071 Valencia,
Spain\\
{\rm and:}  Departament d'Estructura i Constituents de la Mat\`eria,
and Institut de Ci\`encies del Cosmos, Universitat de Barcelona \\
  Diagonal 647, E-08028 Barcelona, Spain\\
}
  
\author{Manuel Pe\~na\footnote{mapeji@ific.uv.es}  and Nuria Rius \footnote{nuria@ific.uv.es}\\
Depto.\ de F\'{\i}sica Te\'orica and
IFIC, Universidad de
Valencia-CSIC \\ 
Edificio de Institutos de Paterna, Apt. 22085, 46071 Valencia,
Spain\\}
\abstract{We analyze leptogenesis in the context of seesaw models with 
almost conserved lepton number, focusing on the $L$-conserving
contribution to the flavoured CP asymmetries. We find that,
contrary to previous claims,  successful leptogenesis is feasible for 
masses  of the lightest heavy neutrino  as low as 
$M_1 \sim 10^6$ GeV, without resorting to the resonant
enhancement of the CP asymmetry for strongly 
degenerate heavy neutrinos. This lower limit renders 
thermal leptogenesis compatible with 
the gravitino bound in supersymmetric scenarios.}
\preprint{%
  IFIC/12-25\\
  FTUV-12-0419\\
ICCUB-12-129}

\begin{document}
\maketitle

\section{Introduction}

Two of the main evidences of physics beyond the Standard Model (SM), namely 
the observed (tiny) neutrino masses  and  the baryon asymmetry of the universe 
(BAU), 
are naturally explained in the context of the  seesaw mechanism 
\cite{Minkowski:1977sc,GellMann:1980vs,Yanagida:1979as,Mohapatra:1979ia}.
In type I seesaw models, the SM is extended with at least two singlet 
Majorana neutrinos 
which can produce the observed BAU via leptogenesis \cite{fukugita86,review}:  
a lepton asymmetry is dynamically generated in 
the out of equilibrium decay of the heavy Majorana neutrinos, and 
then partially converted into a baryon asymmetry due to  
($B+L$)-violating non-perturbative sphaleron interactions~\cite{kuzmin85}.

For a very hierarchical spectrum of heavy singlet neutrinos $M_1 \ll M_2 \ll M_3$, 
the $L$-violating CP asymmetry generated in the decay of the lightest singlet 
has an upper bound proportional to $M_1$, the so-called  
Davidson-Ibarra (DI) bound \cite{davidson02}. 
This implies a lower bound $\sim 10^9$~GeV for the mass of the sterile neutrinos
in order for $N_1$-dominated leptogenesis to be successful.
Careful numerical studies show that the DI bound can be evaded for moderate 
hierarchies, e.g. the lower bound on $M_1$ is relaxed by more than one order
of magnitude with respect to the hierarchical limit one for $M_3/M_2 \sim
M_2/M_1 \sim 10$~\cite{hambye03}. 
However  to reach these low values of $M_1$ some unlikely 
cancellations are needed, which are not motivated by 
any underlying symmetry. 
 Flavour effects~\cite{barbieri99,endoh03,abada06,nardi06,abada06II,blanchet06} do not substantially change 
 this result.
The authors of~\cite{blanchet08} make an analysis of the parameter space for
successful leptogenesis relaxing also the condition of hierarchical neutrinos and
including the $L$-conserving part of the flavoured CP asymmetries, finding  
a lower bound $M_1 \gtrsim 10^8$ GeV.

Such lower bound on $M_1$ in turn yields a lower bound
for the reheating temperature, $T_{RH}$,  of the same order, 
since to thermally produce the neutrinos 
$M_1 \lesssim 5 \,  T_{RH}$ \cite{buchmuller04, giudice04, racker08}.
On the other hand,  in supersymmetric scenarios
gravitinos are copiously produced in the high temperature plasma, 
and their late decay can jeopardize successful 
nucleosynthesis (BBN), leading to 
an upper bound on $T_{RH}$,  which depends on the 
gravitino mass  \cite{Khlopov84,Ellis84}.
If the gravitino is unstable, 
for gravitino masses, $m_{3/2}$, in the natural range from 100 GeV to 1 TeV, and within the 
minimal supergravity framework,  $T_{RH}$  should be smaller than $10^5 - 10^7$ GeV,
while for $m_{3/2} \gtrsim 10$ TeV, $T_{RH}$ can be of order 
$10^9 - 10^{10}$ GeV \cite{Kawasaki08}.
As a consequence, in supersymmetric thermal leptogenesis there is 
some conflict  between the gravitino bound on the reheat temperature
and the thermal production of heavy neutrinos.

There are several 
possible ways out  of this conflict: for instance, if the gravitino is
 stable, the nucleosynthesis bound depends
on the next-to-lightest supersymmetric particle, but $T_{RH} \gtrsim 10^9$ GeV 
can be obtained for $m_{3/2} \gtrsim 10$ GeV \cite{Kawasaki08}.
Also, one can avoid the DI bound resorting to resonant leptogenesis, i.e., a resonant 
enhancement of the CP asymmetry which occurs when there are at least  two strongly 
degenerated heavy neutrinos, such that $M_2 - M_1 \sim \Gamma_N$, being 
$\Gamma_N$ their decay width  \cite{covi96II,Anisimov05}. In this scenario, 
leptogenesis is feasible at much lower temperatures, $T \sim {\cal O}(1 \;\rm{TeV})$
\cite{pilaftsis04,piu,pilaftsis08,pilaftsisnew}. 

Although such quasi-degeneracy of the heavy neutrinos might seem unnatural, 
there are well motivated seesaw models which yield a heavy neutrino quasi-degenerate spectrum, namely those 
with an approximately conserved $B-L$ \cite{mohapatra86,Hernandez09}.
In these models 
 the tiny neutrino masses are proportional to  small 
lepton number-breaking parameters, which are technically natural 
since a larger symmetry is realized when they vanish  \cite{tHooft}.
Moreover, the heavy neutrinos can be much lighter than in the generic seesaw, within the energy 
reach of LHC,  and there can be a large active-sterile neutrino mixing.
Also, lepton flavour violating rare decays as well as non-unitarity of the leptonic mixing matrix
are present even in the limit of conserved $B-L$, and therefore they are unsuppressed
by the light neutrino masses \cite{Bernabeu87,GonzalezGarcia91,Hernandez09}.
As a consequence, 
much attention has been devoted recently to this class of low scale seesaw models,  
since they have a rich phenomenology both at  LHC  \cite{Han06,delaguila07,Kersten07}
and at  low energy charged lepton rare decay experiments, such as 
$\mu \rightarrow e \gamma$, and also lead to successful resonant leptogenesis 
\cite{asaka08,blanchet09,blanchet10}.

In order to see more clearly the structure of seesaw models with small 
violation of $B-L$,
it is useful to expand the low energy effective 
Lagrangian as
\be
\label{Leff} 
\mathcal{L} = \mathcal{L}_{\text{SM}} + \frac{c^{d=5}}{\Lambda_{LN}} 
\mathcal{O}^{d=5} + \sum_i \frac{c_i^{d=6}}{\Lambda^2_{FL}} 
\mathcal{O}_i^{d=6}  + \ldots \ , 
\ee
where the dimensionless couplings $c^{d=5}, c_i^{d=6}$ are assumed to be 
of $\mathcal{O}(1)$.

The only operator of $d=5$  is  Weinberg's operator, responsible 
for neutrino masses, which is suppressed by the lepton number violating scale $\Lambda_{LN}$,
while  flavour changing but lepton number conserving $d=6$ 
operators $\mathcal{O}_i^{d=6}$  are suppressed by a different scale $\Lambda_{FL}$. 
In models with approximately conserved lepton number, 
there is a separation of scales:  $\Lambda_{FL}
\sim \mathcal{O}({\rm TeV})$ can be related to the heavy neutrinos mass scale, but
$\Lambda_{LN} \gg \Lambda_{FL}$, since light neutrino masses only appear when 
small $L$-violating perturbations in the Yukawa couplings and/or the singlet neutrino mass 
matrix are introduced; in this scenario, $\Lambda_{LN}$ 
does not correspond to any particle mass. 
 
It has been noticed that 
even if the heavy neutrino that generates the BAU is not quasi-Dirac, 
or the mass splitting is outside the resonant regime, in seesaw models with almost conserved 
$B-L$ the scale of leptogenesis can be lower than in the standard seesaw  
\cite{antusch09}, provided flavour effects are at work. 
This is so because in these models 
leptogenesis can be driven by the $d=6$ operators, which contribute only  to the 
$L$-conserving part of the flavoured  CP-asymmetries 
 and escape the DI bound, 
even if the heavy  neutrinos are hierarchical, because these operators  are not linked to neutrino masses.
The drawback is that the same $d=6$ operators induce large lepton flavour violating (but 
lepton number conserving) processes, which tend to equilibrate 
 the flavoured asymmetries diminishing the total lepton asymmetry
\cite{aristizabal09,fong10}.

 In this paper we reanalyze the possibility of having successful leptogenesis driven by the 
purely flavoured  $L$-conserving contribution to the CP asymmetries, in the context 
of seesaw models with small violation of $B-L$.
In the numerical analysis of \cite{blanchet08} such contribution was included, 
however models with almost conserved $B-L$ (which can lead to larger CP asymmetries) were not considered in detail and the flavour equilibration processes were not taken into account. 
A study of the lower bound on $M_1$ for approximately $B-L$-conserving models was
carried out in~\cite{antusch09}, but their cross sections are different from ours
and the analysis of the parameter space seems to be non-exhaustive.
We have computed the cross sections for the crucial lepton flavour changing 
processes which lead to flavour equilibration,  and we have found important differences with respect to previous results  \cite{piu}. We have also taken into account decays and inverse decays of the 
next-to-lightest  
neutrinos, neglected in  \cite{antusch09}, and thoroughly
scanned  the parameter space to find the lowest  
$M_1$ able to generate the BAU within this framework.
Moreover, we have included a detailed discussion about leptogenesis in 
models with almost conserved $B-L$, which depends on the heavy neutrino spectrum.

The outline of this paper is as follows.
In Section~\ref{sec:models} we discuss  the main features of leptogenesis 
 within the framework of seesaw models with approximately conserved 
 $B-L$.
In Section~\ref{sec:be} we write the set of Boltzmann equations (BE) relevant for leptogenesis 
in this scenario and compute the cross sections for the lepton flavour violating 
(but total lepton number  conserving) processes, which play a crucial role in models 
with null  (or negligible) total CP asymmetry. 
In Section~\ref{sec:results} we perform a detailed analysis of the parameter space which leads 
to successful leptogenesis with only $L$-conserving CP asymmetries, 
and we conclude in 
Section~\ref{sec:conclusions}.

%%%%%%%%%%%%%%%%%%%%%%%%%%%%%%%%%%%%%%%%%%%%%%%%%%%%%%%%%%%%%%%%%%%%%
\section{Leptogenesis in models with small violation of $B-L$}
\label{sec:models}
In this section we want to determine what can be distinct, regarding
leptogenesis, among different models with small violation of $B-L$. 
 Let us call
$N_1$ the SM fermion singlet that is mainly responsible for the
generation of the lepton asymmetry. Additionally we call $N_2$ the SM fermion singlet
 that makes the most important (non resonant) virtual contribution to the CP
asymmetry in the $N_1$ decays.
They have Yukawa interactions with the lepton
doublets of the SM $\ell_\alpha\; (\alpha=e, \mu, \tau)$ described by the Lagrangian
\begin{equation}
\mathcal{L}_Y = - \lambda_{\alpha i}\,{\widetilde h}^\dag\, \overline{P_R N_i} \ell_\alpha 
- \lambda^*_{\alpha i } \overline{\ell}_\alpha P_R N_i {\widetilde h} \; .
\label{eq:lagy}
\end{equation}
If $B-L$ is only slightly violated, then for each $N_i$ the conditions (i) or (ii) described below must be satisfied:
\begin{itemize}
\item[(i)] $N_i$ is a Majorana neutrino with two degrees of freedom, whose Yukawa interactions violate lepton number and therefore the couplings $\lambda_{\alpha i }$ must be small.
This is  analogous to the standard seesaw, so the $N_i$ contribution to the $d=5$ 
Weinberg operator is given by
\be
(c^{d=5}_M)_{\alpha \beta} = \lambda_{\alpha i } \lambda_{\beta i} \ . 
\ee
\item[(ii)] The $N_i$ is a Dirac or quasi Dirac neutrino with four degrees of freedom;
this means that
  there are two Majorana neutrinos $N_{\ih}$ and $N_{\il}$ with masses $M_i + \mu_i$ and $M_i - \mu_i$ respectively.  The parameter $\mu_i \ll M_i$ measures the amount of $B-L$ violation,
  so that if $B-L$ is conserved, $\mu_i =0$ and $N_i= (N_{\ih} + i N_{\il})/\sqrt{2}$ 
  is a Dirac fermion. 
  The Yukawa interactions can be expressed as
\begin{equation}
\mathcal{L}_{Y_{Ni}} = - \lambda_{\alpha i}\,{\widetilde h}^\dag\,
\overline{P_R \frac{N_{\ih} + i N_{\il}}{\sqrt{2}}} \ell_\alpha  -
  \lambda^\prime_{\alpha i}\,{\widetilde h}^\dag\,
\overline{P_R \frac{N_{\ih} - i N_{\il}}{\sqrt{2}}} 
  \ell_\alpha + h.c.,
\end{equation}
where 
$\lambda^\prime_{\alpha i} \ll 1$. 
The terms proportional to $\lambda^\prime_{\alpha i}$ induce lepton number violation even when $\mu_i \to 0$ and hence they are similar in nature to the ones described in (i). Neglecting these terms, the Yukawa couplings of $N_{\ih}$ and $N_{\il}$ are equal to $\lambda_{\alpha \, ih} = \frac{\lambda_{\alpha i}}{\sqrt{2}}$ and $\lambda_{\alpha \, il} = i \frac{\lambda_{\alpha i}}{\sqrt{2}}$, respectively. 
These $\lambda_{\alpha i}$ can be large, because they do not vanish in the $B-L$ 
conserved limit: in the absence of $\mu_i$ and 
$\lambda^\prime_{\alpha i}$, a perturbatively conserved lepton number can be defined, 
by assigning $L_N =1$  to $N_i$, and $L_{\ell_\alpha} =1$ to the SM leptons.

At leading order in the small $B-L$ breaking parameters ($\lambda'_{\alpha i}, \mu_i/M_i$), 
the contribution of a quasi Dirac heavy neutrino to the  Weinberg operator is  
\be
\frac{(c_{QD}^{d=5})_{\alpha \beta}}{\Lambda_{LN}}= 
(\lambda'_{\alpha i} - \frac{\mu_i}{M_i} \lambda_{\alpha i}) \frac{1}{M_i} \lambda_{\beta i}
+ \lambda_{\alpha i} \frac{1}{M_i} (\lambda'_{\beta i} - \frac{\mu_i}{M_i} \lambda_{\beta i}) 
+ \ldots 
\ee
As a result, one can 
reproduce the tiny light neutrino masses with large Yukawa couplings $\lambda_{\alpha i}$ and heavy neutrino 
masses $M_i$ as low as the TeV scale. 
Moreover, the admixture among singlet and doublet $SU(2)$ neutral leptons
(and the corresponding violation of unitarity in the light neutrino sector) is of order
$\lambda_{\alpha i}  v/M_i$ and can be large.

 In this work we will consider the cases 
\begin{itemize}
\item (iia) $\mu_i \ll \Gamma_{N_{\ih}}, \Gamma_{N_{\il}} \; \,$  (Dirac limit), and 
\item (iib) $\Gamma_{N_{\ih}}, \Gamma_{N_{\il}} \; \ll \mu_i \ll M_i \; \,$  (Majorana limit), 
\end{itemize} 
where the decay widths $\Gamma_s$ are given by  
\begin{equation}
\Gamma_{N_{\ih}} = \frac{M_i + \mu_i}{8 \pi} \frac{\left(\lambda^\dag \lambda \right)_{ii}}{2} \approx \frac{M_i - \mu_i}{8 \pi} \frac{\left(\lambda^\dag \lambda \right)_{ii}}{2} = \Gamma_{N_{\il}} \; .
\end{equation}
Instead we leave the study of $\mu_i \sim \Gamma_{N_{ih, il}}$ for future research because this case cannot be described well with the simple BE used here~\cite{desimone07,garny09,garbrecht11,garny11}.\end{itemize}
Then, the study of leptogenesis for a large class of models with small violation of $B-L$ can be covered by considering the different combinations of options for both $N_1$ and $N_2$. Next we comment on some key points for the different possibilities, while the complete quantitative analysis is made in Sec.~\ref{sec:results}. We assume that $N_1$ is lighter than $N_2$, leaving some remarks of the opposite case $M_2 < M_1$ for Sec.~\ref{sec:results}.
\begin{itemize}
\item[I.] $N_1$ and $N_2$ satisfy (i): This case is not very interesting since the CP asymmetries $\epsilon_{\alpha 1}$ in the decays of $N_1$ into leptons of flavour $\alpha$, $\epsilon_{\alpha 1} \equiv \tfrac{\Gamma(\proname{N_1}{\ell_\alpha h}) - \Gamma(\proname{N_1}{\bar \ell_\alpha \bar h})}{\sum_\alpha \Gamma(\proname{N_1}{\ell_\alpha h}) + \Gamma(\proname{N_1}{\bar \ell_\alpha \bar h})}$, being proportional to the square of the Yukawa couplings of $N_2$, are very small, so it is not different from the standard seesaw.
\item[II.] $N_1$ satisfies (i) and $N_2$ (ii): The $L$-violating part of the CP asymmetry, $\epsilon_{\alpha 1}^{\mislash{L}}$, is suppressed by $\mu_2$ while the $L$-conserving part, $\epsilon_{\alpha 1}^{L}$, is not. To see this, let us write $\epsilon_{\alpha 1} = \epsilon_{\alpha 1}^{\mislash{L}} + \epsilon_{\alpha 1}^{L}$ as~\cite{covi96,roulet97}
\begin{equation}
\epsilon_ {\alpha 1} =  
\sum_{j = \dosh,\dosl} f(a_j) \miim{\lambda_{\alpha j}^* \lambda_{\alpha 1} 
(\lambda^\dag \lambda)_{j 1}} + \sum_{j = \dosh,\dosl} g(a_j) \miim{\lambda_{\alpha j }^* \lambda_{\alpha 1} 
(\lambda^\dag \lambda)_{1 j}} \; ,
\end{equation}
where $f(a_j)$ and $g(a_j)$ are functions of $a_j \equiv M_j^2/M_1^2$ and contain the factor $1/(\lambda^\dag \lambda)_{11}$. The terms proportional to $f(a_j)$ and $g(a_j)$ come from the $L$-violating and $L$-conserving contributions, respectively. To lowest order in $\mu_2$, $f(a_{\dosh}) = f(a_{\dosl})$ and $g(a_{\dosh}) = g(a_{\dosl})$. Taking into account the alignment between the Yukawa couplings of $N_{\dosl}$ and $N_{\dosh}$, $\lambda_{\alpha \, \dosl} = a \lambda_{\alpha \, \dosh}$ with $a=i$, it is clear that
\begin{equation}
\begin{split}
\epsilon_ {\alpha 1} \xrightarrow[\mu_2 \to 0]{} &  f(a_{\dosh}) \miim{\lambda_{\alpha \,\dosh}^* \lambda_{\alpha 1} 
(\lambda^\dag \lambda)_{\dosh \, 1}} (1 + a^{*2}) + \\ & g(a_{\dosh}) \miim{\lambda_{\alpha \, \dosh}^* \lambda_{\alpha 1} 
(\lambda^\dag \lambda)_{1 \, \dosh}} (1 + \abss{a}) \\
& =  0 + 2 g(a_{\dosh}) \miim{\lambda_{\alpha \, \dosh}^* \lambda_{\alpha 1} 
(\lambda^\dag \lambda)_{1 \, \dosh}} \; .
\end{split}
\end{equation}
Hence we see that the contributions from $N_{\dosh}$ and $N_{\dosl}$ add up in the $L$-conserving part and cancel in the $L$-violating one.
 
While the total CP asymmetry $\epsilon_1 \equiv \sum_\alpha \epsilon_ {\alpha 1}$ is
proportional to $\mu_2$ because $\sum_\alpha \epsilon_ {\alpha 1}^{L} = 0$,
the
amount of matter-antimatter asymmetry produced during leptogenesis generally
is not suppressed by $\mu_2$ due to flavour
effects (it is possible to have successful leptogenesis even with
$\epsilon_1=0$~\cite{nardi06}). Nevertheless, this asymmetry does vanish
or becomes proportional to $\mu_2$
when the $L$-violating parameter $(\lambda^\dag \lambda)_{11} \to 0$,
independently of whether the density of $N_1$ at the onset of leptogenesis is
null or thermal. This is because for an initial null density of heavy neutrinos
and small values of $(\lambda^\dag \lambda)_{11}$ the final baryon asymmetry is proportional to $(\lambda^\dag
\lambda)_{11}^2$~\cite{racker08}, while for an initial thermal $N_1$ density
the washouts become negligible and so do the flavour effects, in which case
the total asymmetry generated is proportional to $\epsilon_1 \propto \mu_2$.

\item[III.]  $N_1$ satisfies (ii) and $N_2$ (i): Since the Yukawa couplings of $N_2$
  are small, the virtual contribution of $N_2$ to the CP asymmetries
  $\epsilon_{\alpha \, \unol}$ and  $\epsilon_{\alpha \, \unoh}$ in the decays of
  $N_{\unol}$ and $N_{\unoh}$, respectively, is small. More interesting is the virtual contribution of
  $N_{\unol}$ to $\epsilon_{\alpha \, \unoh}$ and of $N_{\unoh}$ to $\epsilon_{\alpha \, \unol}$, due to the resonant enhancement of the CP asymmetry for degenerate neutrinos. 
  However these contributions are suppressed by $\frac{\lambda^\prime_{\alpha 1}}{\sqrt{(\lambda^\dag \lambda)_{11}}}$ because of  the  alignment of the Yukawa couplings of $N_{\unol}$ and $N_{\unoh}$ for $\lambda^\prime_{\alpha 1}=0$, therefore the CP asymmetry cannot reach the maximum value 1/2 of more generic resonant leptogenesis models. A detailed study of this case was made 
 in~\cite{asaka08}, in the limit $\mu_1 \gg  \Gamma_{N_{\unol}, N_{\unoh}}$: they found that 
it is possible to lower the scale of leptogenesis 
 provided that the parameter which controls 
lepton number violation in the Yukawa couplings $\epsilon= \lambda'/\lambda$ is much larger than 
the one describing the mass splitting between the quasi-Dirac heavy neutrinos, 
$\epsilon_M = \Delta M/M$. For instance, in order to obtain successful leptogenesis 
with $M = 10^6$ GeV (1 TeV), one needs $\epsilon \sim 10^{-3}$ and 
$\epsilon_M \sim 10^{-8}$ ($10^{-11}$). 
 The case $\mu_1 \lesssim \Gamma_{N_{\unol}, N_{\unoh}}$ has been considered in
  \cite{blanchet09}, however it has been shown that in this maximal resonant regime the classical Boltzmann picture breaks down~\cite{desimone07,garny09,garbrecht11,garny11}.

\item[IV.]  $N_1$ and $N_2$ satisfy (ii): 
Besides the resonant contributions which are the same as in III,  
now also the  virtual contribution of $N_2$ to the CP asymmetries
 $\epsilon_{\alpha \, \unol}$ and  $\epsilon_{\alpha \, \unoh}$ can be large, so we focus on 
these terms. 
Neglecting the $\lambda^\prime_{\alpha i} \ll \lambda_{\alpha i}$
couplings, the CP asymmetries in the decays of $N_{\unol}$ and $N_{\unoh}$ become equal. 
Here it is very important to distinguish between the cases 
$\mu_1 \gg \Gamma_{N_{\unol}, N_{\unoh}}$ and $\mu_1 \ll \Gamma_{N_{\unol}, N_{\unoh}}$. 
In the first one, $N_{\unol}$ and $N_{\unoh}$ behave as two independent Majorana neutrinos regarding the generation of the lepton asymmetry, which can hence be roughly double with respect to II. However in the second case $N_1$ is (or effectively behaves as) a Dirac neutrino, i.e. lepton number is conserved in its decay, and therefore the only possibilities to end up with a non-zero baryon asymmetry is to have important washouts from the two Majorana components of 
$N_2$ (if $\mu_2 \gg \Gamma_{N_{\dosl}, N_{\dosh}}$) or let the sphalerons freeze out during leptogenesis~\cite{gonzalezgarcia09}.
\end{itemize}

Summarizing, 
in models with small violation of $B-L$ the CP asymmetries in the decay of 
$N_1$ can be enhanced in cases II, III and IV.  
Since the resonant contribution to the $L$-violating CP asymmetry has been widely studied, 
we develop in Sec.~\ref{sec:results} a quantitative analysis which covers all the 
non-resonant interesting cases, i.e., II and IV with $\mu_1 \gg \Gamma_{N_{\unol}, N_{\unoh}}$.
With respect to $N_2$, we consider two possibilities, 
$\mu_2 \gg \Gamma_{N_{\dosl}, N_{\dosh}}$ and
$\mu_2 \ll \Gamma_{N_{\dosl}, N_{\dosh}}$
(see~\cite{gonzalezgarcia09} for leptogenesis with both $N_1$ and $N_2$ satisfying 
$\mu_i \ll \Gamma_{N_{\il}, N_{ih}}$).

%%%%%%%%%%%%%%%%%%%%%%%%%%%%%%%%%%%%%%%%%%%%%%%%%%%%%%%%%%%%%%%%%%%
\section{Boltzmann equations}
\label{sec:be}
Motivated by the above discussion we consider a scenario for leptogenesis involving three fermion singlets $N_1, N_{\dosl}, N_{\dosh}$ (each of them having two degrees of freedom), with respective masses $M_1, M_2 - \mu_2, M_2 + \mu_2$ and Yukawa couplings given by the Lagrangian
\begin{equation}
\label{eq:lagy12}
\mathcal{L}_{Y} = - \lambda_{\alpha 1}\,{\widetilde h}^\dag\, \overline{P_R N_{1}} \ell_\alpha - \lambda_{\alpha 2}\,{\widetilde h}^\dag\,
\overline{P_R \frac{N_{\dosh} + i N_{\dosl}}{\sqrt{2}}} \ell_\alpha  + h.c. \; .
\end{equation}
Recall that the parameters $\lambda_{\alpha 1}$ violate lepton number and hence $\lambda_{\alpha 1} \ll \lambda_{\alpha 2}$. In turn this implies that the CP asymmetry in $N_1$ decays is the dominant one. As a first approximation, 
in the Eq.~\eqref{eq:lagy12} we have neglected the $L$-violating couplings $\lambda^\prime_{\alpha 2} \ll \lambda_{\alpha 2}$, 
because we expect that  their  contributions to the CP asymmetries and washouts are negligible.
We have checked indeed that this is the case in the parameter space region relevant for leptogenesis that can also accommodate the observed light neutrino masses
(see Sec.~\ref{sec:results}).
As we will explain below, it is convenient to take $M_1 < M_2$ in order to obtain the lowest energy scale for leptogenesis within this framework.

The amount of leptons and antileptons present in the Universe can be described by density matrices in flavour space. The evolution equations of these density matrices take the simplest form in the basis that diagonalize them, which are determined by the fastest interactions in flavour space (if there is a hierarchy among the different interactions). Let us suppose that the fastest interactions during $N_1$-leptogenesis are the $N_2$-Yukawa interactions~\footnote{When the effects of $N_{\dosl}$ and $N_{\dosh}$ simply add up we will refer to these states generically as $N_2$. In the models considered here the only situation which requires a differentiated treatment of the  degrees of freedom associated to $N_2$ is when $\mu_2 \ll \Gamma_{N_{\dosl, \dosh}}$, because a density asymmetry can develop between the states produced by $\ell h$ and $\bar \ell \, \bar h$ (see below).}. 
In this case the BE would be diagonal in a basis $(\ell_2,\ell_{2 \perp},{\ell'}_{\! 2 \perp})$, with $\ell_{2 \perp}$ and ${\ell'}_{\! 2 \perp}$ two orthogonal lepton flavour states perpendicular to the $N_2$-decay eigenstate $\ell_2$. But $\epsilon(\proname{N_1}{\ell_{2 \perp} h,{\ell'}_{\! 2 \perp} h }) = 0$ (with $\epsilon(\proname{N_1}{\ell_a h})$ the CP asymmetry in the decay of $N_1$ into $\ell_a h$), and since the total CP asymmetry is null, then also $\epsilon(\proname{N_1}{\ell_2 h}) = 0$, with the result that no lepton asymmetry would be produced. 
Moreover, no asymmetry is generated when the $N_1$-Yukawa interactions are the fastest ones, because in this case the BE would be diagonal in the basis
$(\ell_1,\ell_{1 \perp},{\ell'}_{\! 1 \perp})$, and since the   
CP asymmetry in the $N_1$-decay eigenstate $\epsilon(\proname{N_1}{\ell_1 h})$ 
coincides with  the total CP asymmetry,  it also vanishes.
So we demand that the couplings of $N_1$ and $N_2$ be small enough, such that the Yukawa interactions of the $\tau$ are the dominant ones. This implies that the BE are diagonal in the orthogonal  basis $(\ell_\tau, \ell_{\tau \perp},{\ell'}_{\! \tau \perp})$, with $\ell_{\tau \perp}$ and ${\ell'}_{\! \tau \perp}$ being determined by the fastest interaction acting in the plane perpendicular to $\ell_\tau$. For simplicity we take the $N_1$ and $N_2$ decay eigenstates to be perpendicular to $\ell_e$, in which case the aforementioned basis is ${\ell_\tau, \ell_\mu, \ell_e}$. Since no asymmetry is generated in the $\ell_e$ flavour, two BE will be enough to determine the evolution of the lepton asymmetry and consequently the flavour indices $\alpha$ and $\beta$ run over the species $\mu$ and $\tau$. Later we will comment on the more general case with three non-null flavour asymmetries.

Besides the charged lepton Yukawa interactions, the most relevant processes are the decays and inverse decays of the different heavy neutrinos, and the $L$-conserving but
$L_\alpha$-violating scatterings $\proname{\ell_\beta h}{\ell_\alpha h}$, 
$\proname{\ell_\beta \bar{h}}{\ell_\alpha \bar{h}}$, and
$\proname{h\bar{h}}{\ell_\alpha\bar{\ell_\beta}}$, hereafter called generically
flavour changing interactions (FCI). We do not consider finite temperature corrections to the particle masses and couplings~\cite{giudice04}, moreover we also neglect spectator processes during leptogenesis and the asymmetry developed among the degrees of freedom of the Higgs~\cite{buchmuller01,nardi05}, as well as $\Delta L=1$ scatterings~\cite{nardi07II,fong10II}~\footnote{We have checked that the inclusion of spectator processes and $\Delta L=1$ scatterings modifies the results by at most a few tens of percent.}. The relevant set of BE for the case $\mu_2 \gg \Gamma_{N_{\dosl, \dosh}}$ is
 \begin{eqnarray}
\frac{\dif Y_{N_1}}{\dif z} &=&\frac{-1}{sHz}
\left(\frac{Y_{N_1}}{Y_{N_1}^{eq}}-1\right) \gamma_{D_1}   \; ,  
\label{eq:beM1}\\
\frac{\dif Y_{\Delta_\alpha}}{\dif z} & =& \frac{-1}{sHz} \left\{
 \left( \frac{Y_{N_1}}{Y_{N_1}^{eq}} - 1
\right)\epsilon_{\alpha 1} \, \gamma_{D_1} - \sum_i \g{N_i}{\ell_\alpha h} y_{\ell_\alpha} \right. \nonumber\\ 
&& \left.  - \sum_{\beta \neq
\alpha} \left( 
\gp{\ell_\beta h}{\ell_\alpha h} + 
\g{\ell_\beta \bar{h}}{\ell_\alpha \bar{h}} + 
\g{h\bar{h}}{\ell_\alpha\bar{\ell_\beta}} 
\right) [y_{\ell_\alpha} - y_{\ell_\beta}] \right\} \; ,
\label{eq:beM2}
\end{eqnarray} 
where $z \equiv M_1/T$ (with $T$ the temperature), $H$ is the Hubble expansion rate, $Y_X \equiv n_X / s$ is the number density 
of a single degree of freedom of the particle species 
$X$ normalized to the entropy density $s$, $y_X \equiv (Y_X - Y_{\bar X})/Y_X^{eq}$ 
is the asymmetry density
normalized to the equilibrium density $Y_X^{eq}$, 
%$Y^{eq}$ is the equilibrium density for any relativistic particle, 
and $Y_{\Delta_\alpha} \equiv Y_B/3 - Y_{L_\alpha}$, 
with $Y_B$ the baryon asymmetry and 
$Y_{L_\alpha}$ the lepton asymmetry in the flavour $\alpha$. Since we are neglecting spectator processes during leptogenesis, the asymmetry in the lepton doublets $\ell_\alpha$ can be expressed in terms of $Y_{\Delta_\alpha}$ using the simple relation  $Y_{\Delta_\alpha} = -Y_{L_\alpha}= -2y_{\ell_\alpha} Y_{\ell_\alpha}^{eq}$~\cite{nardi05}. The sphaleron processes convert part of the final $B-L$ asymmetry ($Y_{B-L}^f = \sum_\alpha Y_{\Delta_\alpha}^f$) into a baryon asymmetry ($Y_B^f$) and the relation between these quantities is taken to be $Y_B^f = 28/79 \, Y_{B-L}^f$~\cite{harvey90}. In Eq.~\eqref{eq:beM2} we have also introduced the notation $\g{a, b, \dots}{c, d, \dots}
\equiv \gamma(\proname{a, b, \dots}{c, d, \dots})$ for the reaction density of the process $\proname{a, b, \dots}{c, d, \dots}$, the prime in $\gp{\ell_\beta h}{\ell_\alpha h}$ indicating that
the on-shell contribution has to be subtracted, and $\gamma_{D_i} \equiv \sum_\alpha \g{N_i}{\ell_\alpha h} +
\g{N_i}{\bar{\ell_\alpha} \bar{h}}$. 
Finally notice that the lepton asymmetries generated by $N_2$ decays, $\epsilon_{\alpha 2}$, have not been taken into account, given that $\epsilon_{\alpha 2} \ll \epsilon_{\alpha 1} $.

Instead, if $\mu_2 \ll \Gamma_{N_{\dosl, \dosh}}$ then $N_{\dosl}$ and $N_{\dosh}$ combine to form a Dirac neutrino $N_2 \equiv (N_{\dosh} + i N_{\dosl})/\sqrt{2}$, and therefore there is an asymmetry generated among the  degrees of freedom of $N_2$ which has to be taken into account~\cite{gonzalezgarcia09}. An appropriate set of BE for this case is
 \begin{eqnarray}
\frac{\dif Y_{N_1}}{\dif z} &=&\frac{-1}{sHz}
\left(\frac{Y_{N_1}}{Y_{N_1}^{eq}}-1\right) \gamma_{D_1}   \; ,  
\label{eq:beD1}\\
%\frac{\dif Y_{N_2 + \bar N_2}}{\dif z} &=&\frac{-1}{sHz}
%\left(\frac{Y_{N_2 + \bar N_2}}{Y_{N_2 + \bar
%N_2}^{eq}}-1\right) \gamma_{D_2}\; ,  
%\label{eq:beD2}\\
\frac{\dif Y_{N_2 - \bar N_2}}{\dif z} &=& \frac{-1}{sHz} \sum_\alpha \g{N_2}{\ell_\alpha h} \left[ y_{N_2} - y_{\ell_\alpha} \right] \; ,
\label{eq:beD3}\\
\frac{\dif Y_{\Delta_\alpha}}{\dif z} & =& \frac{-1}{sHz} \left\{
 \left( \frac{Y_{N_1}}{Y_{N_1}^{eq}} - 1
\right)\epsilon_{\alpha 1} \, \gamma_{D_1}  - \g{N_1}{\ell_\alpha h} y_{\ell_\alpha} + \g{N_2}{\ell_\alpha h} \left[ y_{N_2} -
y_{\ell_\alpha} \right]  \right. \nonumber\\ 
&& \left.  -  \sum_{\beta \neq
\alpha} \left( 
\gp{\ell_\beta h}{\ell_\alpha h} + 
\g{\ell_\beta \bar{h}}{\ell_\alpha \bar{h}} + 
\g{h\bar{h}}{\ell_\alpha\bar{\ell_\beta}} 
\right) [y_{\ell_\alpha} - y_{\ell_\beta}] \right\} \; .
\label{eq:beD4}
\end{eqnarray} 

The reaction densities for the FCI are obtained integrating the corresponding cross section $\sigma_{\text{FCI}}$,  $\displaystyle \gamma_{\text{FCI}} = \frac{T}{64 \pi^4} \int_0^\infty \dif s \, s^{1/2} \cst_{\text{FCI}} K_1 \left(\frac{\sqrt{s}}{T}\right)$, with the reduced cross sections $\cst_{\text{FCI}} \equiv 2 s \sigma_{\text{FCI}}$. In order to give general expressions for these cross sections we take the FCI to be mediated by any number of Dirac or Majorana neutrinos $N_i \, (i=1,2, \dots)$ with masses $M_i$, whose Yukawa couplings with the lepton doublets $\ell_\alpha$ are called $h_{\alpha i}$. We find~\footnote{Our expressions for these cross sections differ from the ones used in~\cite{antusch09}, which were taken from~\cite{piu}.}:
\begin{eqnarray}
\cst(\proname{\ell_\beta h}{\ell_\alpha h}) &=& \frac{1}{4 \pi} \sum_{i,j} h^*_{\alpha i} h_{\beta i} h_{\alpha j} h^*_{\beta j} \frac{s^2 \big( s - M_i^2 - i M_i \Gamma_i \big) \big( s - M_j^2 + i M_j \Gamma_j\big)}{\left[ \big( s - M_i^2 \big)^2 + M_i^2 \Gamma_i^2 \right] \left[ \big( s - M_j^2 \big)^2 + M_j^2 \Gamma_j^2 \right]} \; ,\nonumber \\
\cst(\proname{\ell_\beta \bar{h}}{\ell_\alpha \bar{h}}) &=& \frac{1}{2 \pi} \sum_{\begin{subarray}{c} i, j\\ i \neq j \end{subarray}} h^*_{\alpha i} h_{\beta i} h_{\alpha j} h^*_{\beta j}  \frac{1}{M_i^2 - M_j^2} \left\{M_i^2 \ln \frac{s+M_i^2}{M_i^2} - M_j^2 \ln \frac{s+M_j^2}{M_j^2}
    \right\} \nonumber \\
& \quad & + \; \frac{1}{2 \pi} \sum_{\begin{subarray}{c} i,j \\ i=j \end{subarray}}
    h^*_{\alpha i} h_{\beta i} h_{\alpha j} h^*_{\beta j} \left\{  \ln
    \frac{s+M_i^2}{M_i^2} - \frac{s}{s+M_i^2}
    \right\}\; ,\nonumber\\
\cst(\proname{h\bar{h}}{\ell_\alpha\bar{\ell_\beta}}) &=& \frac{1}{2 \pi} \sum_{\begin{subarray}{c} i,j \\ i \neq j \end{subarray}}  h^*_{\alpha i} h_{\beta i} h_{\alpha j} h^*_{\beta j} \left\{ -1 + \frac{M_i^2 (s+M_i^2)}{s(M_i^2 - M_j^2)} \ln
    \frac{s+M_i^2}{M_i^2} + \right. \nonumber \\ &\quad& \hspace{5cm} \left. \frac{M_j^2 (s+M_j^2)}{s(M_j^2 - M_i^2)} \ln
    \frac{s+M_j^2}{M_j^2}   \right\}+ \nonumber\\
& \quad &\; \frac{1}{2 \pi} \sum_{\begin{subarray}{c} i,j\\ i=j \end{subarray}}
    h^*_{\alpha i} h_{\beta i} h_{\alpha j} h^*_{\beta j} \left\{  -2 + \frac{s+2M_i^2}{s} \ln
    \frac{s+M_i^2}{M_i^2} \right\}\;. 
\end{eqnarray}
Moreover, the subtracted reaction density is given by
\begin{equation}
\gp{\ell_\beta h}{\ell_\alpha h} = \g{\ell_\beta h}{\ell_\alpha h} - \sum_i \g{\ell_\beta h}{N_i} \mibr{N_i}{\ell_\alpha h} \; ,
\end{equation}
with $\mibr{N_i}{\ell_\alpha h}$ the branching ratio of the process $N_i \,\rightarrow \, \ell_\alpha h$.

For the models we are considering the dominant contribution to the FCI comes from the exchange of a neutrino with mass $M_2$ and Yukawa couplings $h_{\alpha 2} = \lambda_{\alpha 2}$~\footnote{Note that the contribution to the FCI of the two Majorana neutrinos 
$N_{\dosl}, N_{\dosh}$ with the Yukawa couplings of Eq.~\eqref{eq:lagy12}
is the same as the contribution of one
Dirac neutrino with mass $M_2$ and Yukawa couplings $\lambda_{\alpha 2}$, in the limit $\mu_2 \ll M_2$.}.

%%%%%%%%%%%%%%%%%%%%%%%%%%%%%%%%%%%%%%%%%%%%%%%%%%%%%%%%%%%%%%%%%%%
\section{Results}
\label{sec:results}
We will find the lower bound for $M_1$ under the requirement that
leptogenesis be successful in these scenarios.
The relevant parameters for leptogenesis are $M_1$, $M_2/M_1$, $(\lambda^\dag \lambda)_{11}$, $(\lambda^\dag \lambda)_{22}$, the projectors $K_{\alpha i} \equiv \lambda_{\alpha i} \lambda_{\alpha i}^* /(\lambda^{\dagger} \lambda)_{ii}$ $(\alpha=\mu, \tau; i=1,2),$ and $\mu_2$. Next we comment
on the role of these parameters.
\begin{itemize}
\item {\bf $M_1$}: For a fix value of the Yukawa couplings (and in particular
  of the CP asymmetry when $M_2/M_1$ is kept fixed) the lower is  $M_1$, the stronger the washouts
  become (because the expansion rate becomes slower with decreasing $T$). This is why $M_1$ cannot be very low.
\item {\bf $M_2/M_1$}: The hierarchy among the masses appears in the CP
  asymmetry, with $\epsilon_{\alpha 1} \propto (M_2/M_1)^{-2}$ for $M_2 \gg M_1$, and in the FCI when these are mediated
  mainly by $N_2$ (as always happens in the interesting cases for this work), $\gfe(T) \propto (M_2/M_1)^{-4}$ for $T \sim M_1 \ll
  M_2$. Also note that if $M_2/M_1 \lesssim 20$ the inverse decays of
  $N_2$ could erase part of the asymmetry generated during $N_1$ leptogenesis
  and therefore they should be included in the BE. 
\item {\bf $(\lambda^\dag \lambda)_{11}$}: The intensity of the washouts due to processes
  involving $N_1$ is determined by the effective mass $\tilde m_1 \equiv (\lambda^\dag \lambda)_{11} v^2/M_1$. In order to make full use of the -mandatory- flavour effects we take $\tilde m_1 \gtrsim  m_* \simeq 10^{-3}$~eV~\footnote{The
quantity $m_*$ is the {\it equilibrium mass}, which is defined by the
condition $\tfrac{\Gamma_{N_1}}{H(T=M_1)}=\frac{\tilde m_1}{m_*}$, so that $m_*=\tfrac{16}{3
\sqrt{5}} \pi^{5/2} \sqrt{g_{*SM}} \frac{v^2}{m_{pl}} \simeq 1,08
\times 10^{-3}$~eV ($g_{*SM}$ is the number of SM
relativistic degrees of freedom at temperature $T$ and $m_{pl}$ is the
Planck mass).}. However, since $(\lambda^\dag \lambda)_{11}$ violates lepton number it cannot be very large in the scenarios we are considering.
\item {\bf $(\lambda^\dag \lambda)_{22}$}: The CP asymmetry $\epsilon_{\alpha 1}$
  is directly proportional to $(\lambda^\dag \lambda)_{22}$, hence this
  parameter should be taken as large as possible. There are two reasons that prevent it from being very large. One is that the washouts due to processes
  involving $N_2$ increase with $(\lambda^\dag \lambda)_{22}$. The other is that, as explained at the beginning of Sec.~\ref{sec:be}, the Yukawa interactions of $N_2$ must be slower than those of the $\tau$. Below we explain in more detail how this constraint has been dealt with.

\item {\bf $K_{\alpha i}$}:
The flavoured CP asymmetries depend on the square root of the projectors while the washouts in the flavour ``$\alpha$'' depend linearly on $K_{\alpha i}$, hence the washouts decrease faster than the CP asymmetries with decreasing projectors. This fact must be taken into account in order to maximize the production of lepton asymmetry. Note that since $\sum_\alpha K_{\alpha i} = 1$ and we are taking $K_{ei} = 0$, there are only two independent projectors, which can be chosen as $K_{\mu 1}$ and $K_{\mu 2}$.

\item {\bf $\mu_2$}: 
Since we are not considering the case $\mu_2 \sim \Gamma_{N_{\dosl, \dosh}}$, $\mu_2$ only enters as a discrete parameter, the baryon asymmetry taking one $\mu_2$-independent value when $\mu_2 \gg \Gamma_{N_{\dosl, \dosh}}$ and another when $\mu_2 \ll \Gamma_{N_{\dosl, \dosh}}$. 

\end{itemize}

\begin{figure}[!htb]
\centerline{\protect\hbox{
\epsfig{file=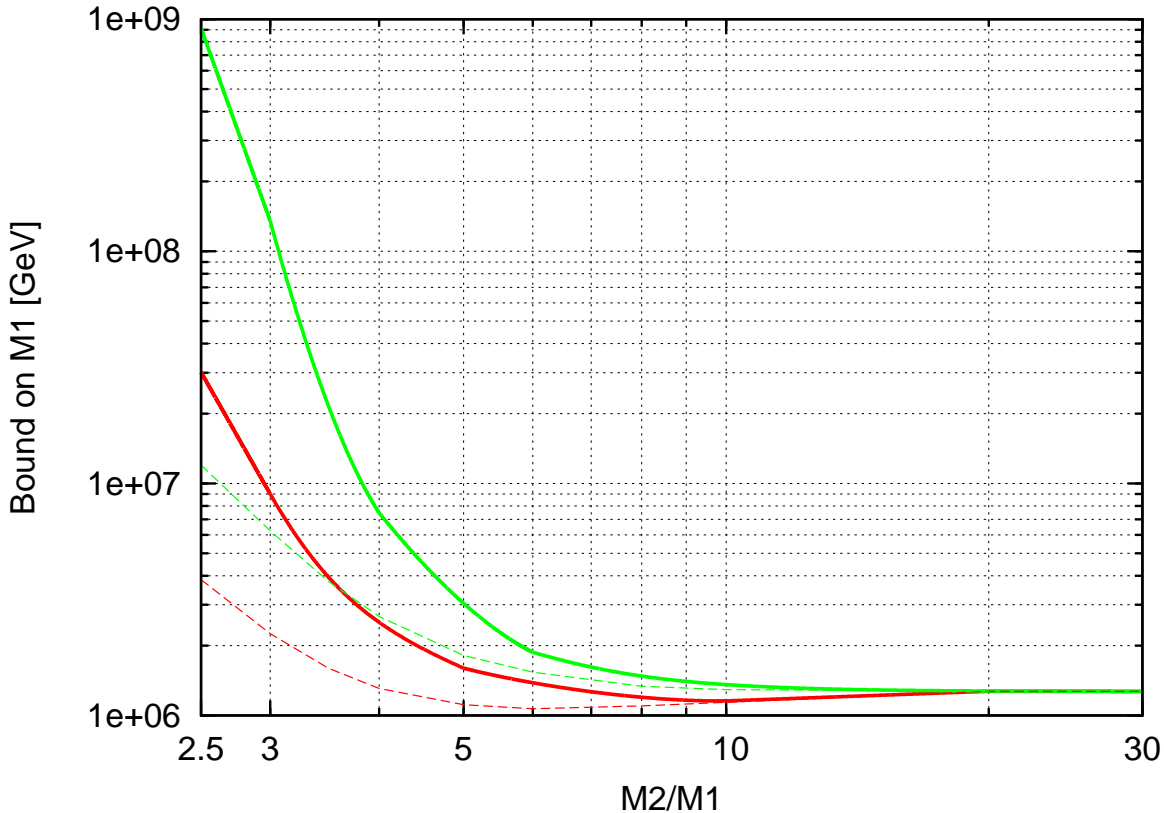 ,width=0.5\textheight,angle=270}}}
\caption[]{Lowest value of $M_1$ yielding successful leptogenesis as a function of $M_2/M_1$. The red curves are for the case $\mu_2 \gg \Gamma_{N_{\dosl, \dosh}}$ and the green ones for $\mu_2 \ll \Gamma_{N_{\dosl, \dosh}}$. The thick continuous curves give the physically correct bound, while the thin dashed ones show the result that would be obtained if the Yukawa couplings of $N_2$ were allowed to take values as large as 1 for all values of $M_2/M_1$.} 
\label{fig:1}
\end{figure}

Given that the hierarchy $M_2/M_1$ is an interesting and crucial parameter, we have determined the minimum value of $M_1$ compatible with successful leptogenesis as a function of $M_2/M_1$, maximizing the final baryon asymmetry over the remaining parameters, i.e. over the relevant combinations of Yukawa couplings: $(\lambda^\dag \lambda)_{11}, (\lambda^\dag \lambda)_{22}, K_{\mu 1}$, and $K_{\mu 2}$. To obtain the baryon asymmetry we have solved numerically the appropriate set of BE, and to get successful leptogenesis we have required
$Y_B  = 8.75 \times 10^{-11}$ \cite{komatsu10}.
The result is represented with the thick continuous curves in Fig.~\ref{fig:1}, the red line corresponding to the case $\mu_2 \gg \Gamma_{N_{\dosl, \dosh}}$ and the green one to $\mu_2 \ll \Gamma_{N_{\dosl, \dosh}}$. 

As can be seen it is possible to have neutrino masses as low as $M_1 \sim 10^{6}$~GeV, i.e. around three orders of magnitude below the lower bound for the standard case of type I seesaw with hierarchical heavy neutrinos. An interesting consequence is that this value can be compatible with the upper bound on the reheating temperature required to avoid the gravitino problem in SUGRA models. Moreover, $M_1$ values around $10^{6}$~GeV can be achieved for a wide range of $N_2$ masses and also for different values of the Yukawa couplings. In particular, the maximum value of the baryon asymmetry is obtained for $\tilde m_1 \sim 10^{-2}$~eV and $K_{\mu 1}$ (or $K_{\tau 1}) \sim 0.1$, but it does not change much if $\tilde m_1$ is varied within the range $5 \times 10^{-3}$ eV $\lesssim \tilde m_1 \lesssim 10^{-1}$ eV as long as the projectors are adjusted in order to have weak washouts in one flavour and strong washouts in the other, e.g. $K_{\mu 1} \tilde m_1 \lesssim m_*$ and  $K_{\tau 1} \tilde m_1 \gtrsim (5-10) m_*$. Regarding the Yukawa couplings of $N_2$, $(\lambda^\dag \lambda)_{22}$ lies approximately between\footnote{In the analysis presented here we have restricted $\abs{\lambda_{\alpha 2}} \le 1$, but we have checked that the results do not change significantly when allowing these Yukawa couplings to take somewhat larger values (but below the perturbative bound).} 0.01 and 1, while for intermediate to large $N_2$ masses, $M_2 \gtrsim 5 M_1$, the smallest projector can take values as large as 0.1 to 0.5, without changing the bound shown in Fig.~\ref{fig:1} by more than a factor 2. Instead, for $M_2 \lesssim 5 M_1$, some hierarchy among the Yukawa couplings of $N_2$ is required, namely 
$\lambda_{\mu 2}/\lambda_{\tau 2}$ (or $\lambda_{\tau 2}/\lambda_{\mu 2}$) 
$\sim (1-3) \times 10^{-2}$ to achieve the values shown in Fig.~\ref{fig:1}.

\begin{figure}[!htb]
\centerline{\protect\hbox{
\epsfig{file=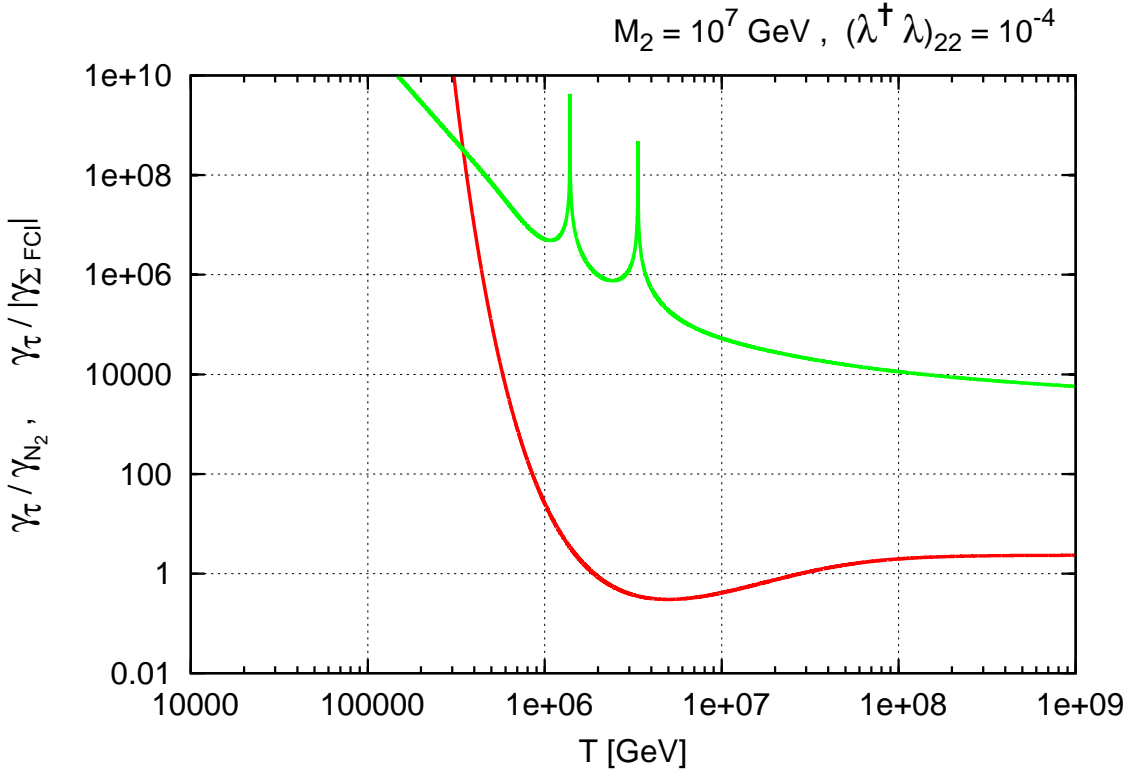 ,width=0.47\textheight,angle=270}}}
\caption[]{Comparison between the rates of the Yukawa interactions of $N_2$ and $\tau$ as a function of the temperature. The red line gives $\gamma_{\tau}/\gamma_{N_2}$ and the green one $\gamma_{\tau}/\abs{\gamma_{\Sigma\,\text{FCI}}}$, where $\gamma_\tau$ is the rate of the Yukawa interactions of the $\tau$, $\gamma_{N_2}$ is the sum of the rates of processes involving a real $N_2$ (here we have included the decay and scatterings with the top quark summing over all lepton flavours), and $\gamma_{\Sigma\, \text{FCI}}$ is basically the sum of the rates of the FCI mediated by an off-shell $N_2$, 
$\gamma_{\Sigma\,  \text{FCI}} \equiv \left( \gp{\ell_\tau h}{\ell_\mu h} + 
\g{\ell_\tau \bar{h}}{\ell_\mu \bar{h}} + 
\g{\ell_\tau \bar{\ell_\mu}}{h\bar{h}}\right)/(K_{\mu 2} K_{\tau 2})$ . We have taken $M_2 = 10^7$~GeV and $(\lambda^\dag \lambda)_{22} = 10^{-4} \simeq h_\tau^2$ (with $h_\tau$ the Yukawa coupling of the $\tau$). Note that the interactions involving a real $N_2$ scale as $(\lambda^\dag \lambda)_{22}$, while the FCI are proportional to $(\lambda^\dag \lambda)_{22}^2$. The corresponding curves for other values of $M_2$ can be obtained simply by making the appropriate translation along the $T$-axis. The spiky shape of the green curve for $0.1 M_2 \lesssim T \lesssim M_2$ is due to the substraction of the on-shell contribution to the process $\proname{\ell_\tau h}{\ell_\mu h}$.} 
\label{fig:2}
\end{figure}

\begin{figure}[!htb]
\centerline{\protect\hbox{
\epsfig{file=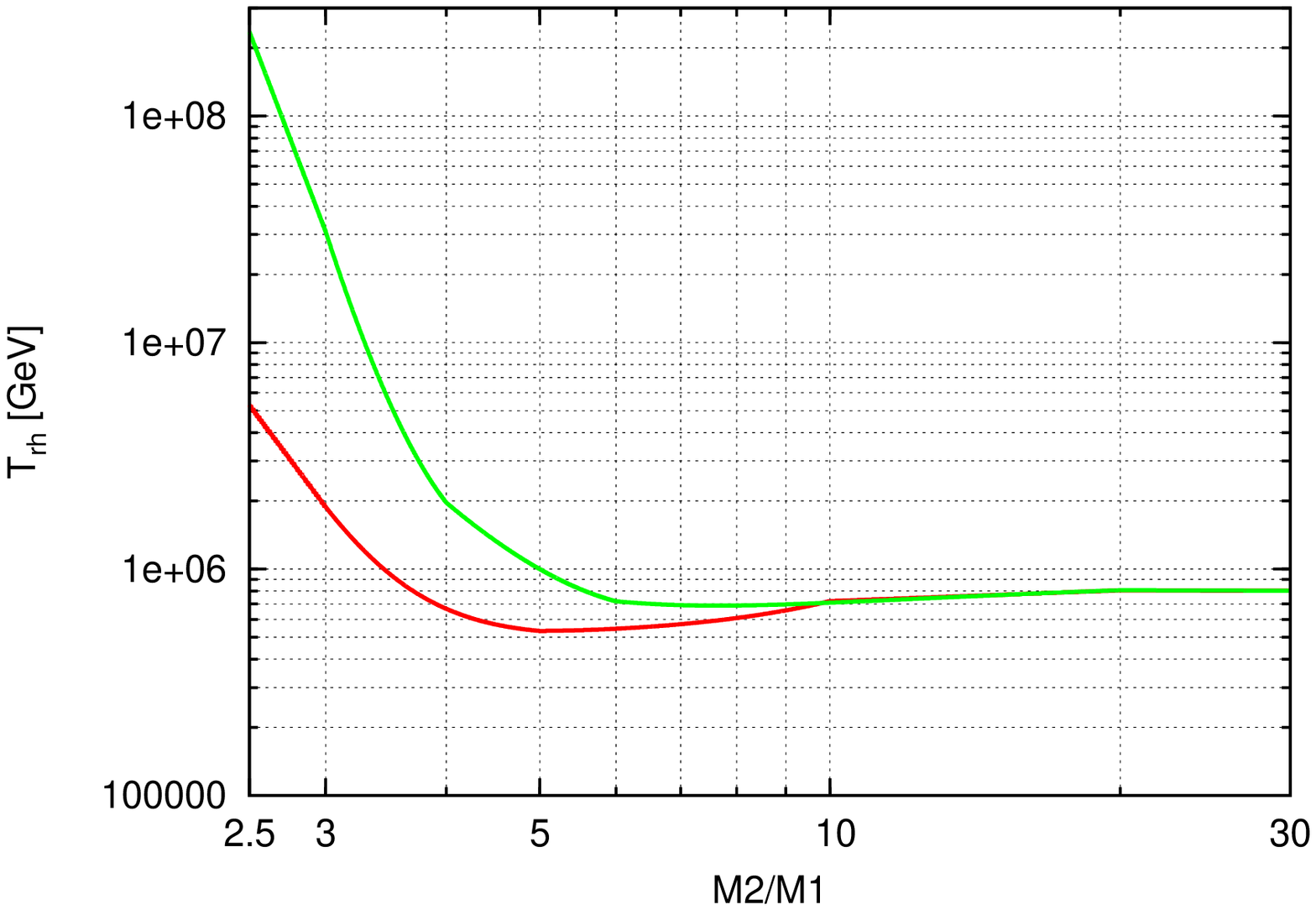 ,width=0.5\textheight,angle=270}}}
\caption[]{The lower bound on the reheating temperature as a function of $M_2/M_1$. The red curve is for the case $\mu_2 \gg \Gamma_{N_{\dosl, \dosh}}$ and the green one for $\mu_2 \ll \Gamma_{N_{\dosl, \dosh}}$.} 
\label{fig:3}
\end{figure}

An important issue for obtaining the bound on $M_1$ is to determine how large the  Yukawa  couplings of $N_2$ can be without violating the condition that the rates of processes involving $N_2$ be slower than the rates of the $\tau$-Yukawa interactions
\footnote{There are not relevant experimental bounds on the Yukawa couplings of $N_2$ for the masses of heavy neutrinos we are considering.}. For reference purposes we plot in Fig.~\ref{fig:2} the ratio between the rates of these interactions, distinguishing between processes with a real or virtual $N_2$, for $M_2 = 10^7~$GeV and $\sqrt{(\lambda^\dag \lambda)_{22}} = 0.01 \sim h_{\tau}$ (the rates of the $\tau$-Yukawa interactions have been taken from~\cite{campbell92,cline93}). On one hand, note that the $\tau$-Yukawa interactions tend to become dominant over the $N_2$ ones as the temperature decreases. On the other hand, for each set of parameters there exists a temperature $T_{in}$ such that the final baryon asymmetry $(Y_B^f)$ does not depend on what happens at $T > T_{in}$ (because there are strong washouts for all the interesting regions of the parameter space). Then the largest possible value for $(\lambda^\dag \lambda)_{22}$ has been set by requiring that the $\tau$-Yukawa interactions be faster than the $N_2$ ones for all temperatures {\it below} $T_{in}$. In practice we have determined $T_{in}$ finding the lowest value of $T$ such that $Y_B^f$ does not change by more than $10\%$ when the initial conditions at $T$, namely $Y_{N_i}(T)$  and $Y_{\Delta_\alpha}(T)$, are varied.  The value of $T_{in}$ as a function of $M_2/M_1$ is shown in Fig.~\ref{fig:3}, reinterpreted as an approximate lower bound for the reheating temperature. Also note that the independence from the initial conditions is an interesting feature by itself, since the models become more predictive.

For comparison we have also plotted in Fig.~\ref{fig:1} the -wrong- bound that would be obtained if $(\lambda^\dag \lambda)_{22}$ were allowed to be as large as 1, ignoring the above discussion. It is clear that as $M_2$ approaches $M_1$ the requirement of an upper bound for $(\lambda^\dag \lambda)_{22}$ becomes very relevant.  This requirement is also the reason why it is convenient to take $M_2 > M_1$, because in the opposite case (when the lepton asymmetry is produced by the decay of the next-to-lightest singlet neutrino and that is usually called ``$N_2$ leptogenesis'') the bound on $(\lambda^\dag \lambda)_{22}$ would not allow for large values of the CP asymmetry.

Another feature apparent in Fig.~\ref{fig:1} is the constant behaviour of the bound on $M_1$ for large values of $M_2/M_1$. This can be understood quite easily as follows. The asymmetry $Y_B^f$ is proportional to the CP asymmetry in $N_1$ decays, $\epsilon_{\mu 1}=-\epsilon_{\tau 1}$, which in turn satisfies $\epsilon_{\mu 1} \propto (M_1/M_2)^2 (\lambda_{\mu 2} \lambda_{\tau 2})$. For $M_2 \gg M_1$, $Y_B^f$ increases with $(\lambda_{\mu 2} \lambda_{\tau 2})$ up to a certain value $(\lambda_{\mu 2} \lambda_{\tau 2})_{\text{max}}$ for which the FCI become important, equilibrating the asymmetries generated in the two flavours with the consequent decrease of the baryon asymmetry~\cite{aristizabal09,fong10}. At the temperatures relevant for leptogenesis, $T \sim M_1 \ll M_2$, the rates of the FCI scale as $\gfe \propto (M_1/M_2)^4 (\lambda_{\mu 2} \lambda_{\tau 2})^2$. Hence we see that if $(M_1/M_2)$ is decreased by a factor $a$, $(\lambda_{\mu 2} \lambda_{\tau 2})_{\text{max}}$ can be increased by a factor $a^2$ keeping constant the rates of the FCI. Therefore $(\lambda_{\mu 2} \lambda_{\tau 2})_{\text{max}} \propto (M_1/M_2)^{-2}$ and hence the maximum value of  $Y_B^f$, being proportional to $(M_1/M_2)^2 \, (\lambda_{\mu 2} \lambda_{\tau 2})_{\text{max}}$, becomes independent of $M_2/M_1$.

As mentioned in Sec.~\ref{sec:be}, for simplicity we have taken $\ell_e$ to be perpendicular to the decay eigenstates of $N_1$ and $N_2$, so that only two flavour asymmetries are generated. We have checked that in the more general three flavour case it is possible to lower the bound on $M_1$ by a factor up to almost 4 with respect to the two flavour case. This is due to the combination of two effects. On one hand, $Y_B^f$ can be almost twice as large compared to the two flavour case (this happens when $K_{\alpha 1} \tilde m_1 \lesssim m_*$ for two different flavours). On the other hand, it can be shown that for a given value of $M_2/M_1$, the maximum value of $Y_B^f$ is proportional to $\sim \sqrt{M_1}$ in the relevant region of the parameter space. Hence an increase in $Y_B^f$ by a factor 2 leads to a decrease in the lower bound of $M_1$ by a factor 4.

Finally let us comment on the relation between the parameters defined above and the light neutrino masses. To lowest order in the $L$-violating parameters, the light neutrino masses $m_{\nu}$ are  given by 
\begin{equation}
\label{eq:numass}
(m_{\nu})_{\alpha \beta}  
\sim
\lambda_{\alpha 1} \frac{v^2}{M_1} \lambda_{\beta 1}
+
(\lambda'_{\alpha 2} - \frac{\mu_2}{M_2} \lambda_{\alpha 2}) \frac{v^2}{M_2} \lambda_{\beta 2}
+ \lambda_{\alpha 2} \frac{v^2}{M_2} (\lambda'_{\beta 2} - \frac{\mu_2}{M_2} \lambda_{\beta 2}) 
\end{equation}
where $v = \langle h \rangle = 174$ GeV is the vev of the Higgs field. 

We have seen that $M_1$ is minimized for values of 
$\tilde m_1 \equiv (\lambda^\dag \lambda)_{11} v^2/M_1$ in the range 
$10^{-2}$ eV $\lesssim \tilde m_1 \lesssim 10^{-1}$ eV,
therefore   the contribution of $N_1$  to light neutrino masses is expected to be of the same order, barring cancellations due to phases. Imposing that such contributions are 
of order $m_{atm} \sim 0.05$~eV, we get $\lambda_{\alpha 1} \sim 10^{-5} - 10^{-4}$. 
 
To reproduce the observed solar and atmospheric mass scales, at least one of the 
contributions from $N_2$ in \eqref{eq:numass} should be also of order $10^{-2}$ eV;
for the parameters that minimize $M_1$, this requirement leads to
$\mu_2/M_2 \sim 10^{-8} - 10^{-6}$, independently of the ratio $M_2/M_1$.  
Moreover, $\Gamma_{N_{\dosl, \dosh}}/M_2 \sim 5 \times (10^{-4} - 10^{-2})$, therefore typically 
$\mu_2 \ll \Gamma_{N_{\dosl, \dosh}}$. However, for $M_1 \gtrsim 5 \times 10^{6}$~GeV, and still not considering large fine tunings related to phase cancellations, smaller values of 
$\lambda_{\alpha 2}$ can lead to successful leptogenesis, and in this region it is possible to have $\mu_2 \gtrsim \Gamma_{N_{\dosl, \dosh}}$.    

With respect to the $L$-violating parameters $\lambda^\prime_{\alpha 2}$, 
their contribution to the masses of the light neutrinos is  $m_{\nu} \sim m_{atm}$ 
 typically for $\lambda^\prime_{\alpha 2} \sim 10^{-8} - 10^{-7}$. We have checked for consistency that  these small values give a negligible 
 contribution to leptogenesis (more specifically to the CP asymmetries and washouts).

%%%%%%%%%%%%%%%%%%%%%%%%%%%%%%%%%%%%%%%%%%%%%%%%%%%%%%%%%%%%%%%%%%
\section{Conclusions}
\label{sec:conclusions}

We have studied leptogenesis in the framework of the seesaw mechanism with small violation of $B-L$. If $B-L$ is only slightly broken, then either the heavy  neutrinos which generate the 
baryon asymmetry  are almost degenerate  and combine to 
form quasi-Dirac fermions, which can have  large lepton number conserving Yukawa 
couplings, or 
 they are Majorana fermions with small lepton number violating Yukawa couplings.

In both cases there are interesting consequences for leptogenesis: in the first one, 
 the strong degeneracy of the heavy neutrinos leads to a 
 resonant enhancement of the CP asymmetry,
 avoiding the DI bound on $M_1$ which applies to
 hierarchical SM singlets. 
 In the second case, the  $L$-conserving part of the flavoured CP
asymmetries in $N_1$ decays 
can be much larger than the $L$-violating one, since the 
former is not linked to light neutrino masses and also escapes the DI bound, 
even if the heavy neutrinos are hierarchical.
As a consequence, in models with almost conserved $B-L$ successful leptogenesis may be 
possible at lower  temperatures than in the standard seesaw, alleviating the 
gravitino problem in supersymmetric  scenarios.

In this paper we have focused on  the second possibility, i.e., we have not
 resorted to the resonant enhancement of the CP asymmetry.
Since the  $L$-conserving contributions to the flavoured CP
asymmetries cancel in the total CP asymmetry,
$\epsilon_i = \sum_\alpha \epsilon_{i \alpha}$,  it is mandatory that flavour effects 
are at work for these terms to have an impact in leptogenesis.

We have exhaustively scanned the parameter space of seesaw models with almost conserved 
$B-L$,  in which the $L$-conserving piece of the flavoured CP asymmetries 
dominates (purely flavoured leptogenesis).
We have found that the largest 
baryon asymmetry is generated by the lightest SM singlet, 
and it rapidly decreases if $|M_2 - M_1| \lesssim 2 M_1$ (but far from 
the resonance region,  $|M_2 - M_1| \sim \Gamma_2$),
being $N_2$ the next-to-lightest heavy neutrino;
thus we have restricted the heavy neutrino masses to the region 
$M_2 \gtrsim 2 M_1$.
The relevant parameters are $M_1$, $M_2/M_1$, 
$(\lambda^{\dagger} \lambda)_{11}$, 
$(\lambda^{\dagger} \lambda)_{22}$ and the flavour projectors 
$K_{\alpha i}$. 
The same $N_2$-Yukawa couplings which enhance the $L$-conserving CP 
asymmetries induce large FCI mediated by $N_2$, 
which tend to equilibrate the asymmetries in the different lepton flavours diminishing the total lepton asymmetry, 
especially  if both $N_1$ and $N_2$ have similar 
masses and are simultaneously present in the thermal bath. 
Therefore, for each value of $M_2/M_1$
we have determined the minimum  $M_1$
compatible with successful leptogenesis, maximizing the final baryon asymmetry over the 
remaining parameters.
We have solved numerically the relevant set of BE, including 
decays and inverse decays of the two singlet neutrino species, as well as 
the FCI. 
We have considered two possibilities: $N_2$ is a pseudo-Dirac fermion
(i.e., two Majorana neutrinos with masses $M_2 \pm \mu_2$, $\mu_2 \ll M_2$) and 
$N_2$ is approximately Dirac, in which case an asymmetry between $N_2$ and
$\bar{N}_2$ is generated and should be taken into account.  
In both cases we have found that leptogenesis is possible for 
$M_1 \gtrsim 10^6 \,$ GeV, as long as $M_2/M_1 \gtrsim 5$
and $(\lambda^{\dagger} \lambda)_{22} \sim 0.01 - 1$
(see Fig.~\ref{fig:1}). 
So purely flavoured leptogenesis in seesaw models with slightly broken $B-L$ provides a solution to the conflict between the upper bound on $T_{RH}$ required to solve the gravitino problem of supersymmetric scenarios and the lower bound on $T_{RH}$ needed for successful thermal leptogenesis. 
However, such heavy neutrinos are far outside the reach of present and near future
colliders and do not lead to observable lepton flavour violation in non-supersymmetric 
frameworks.

%%%%%%%%%%%%%%%%%%%%%%%%%%%%%%%%%%%%%%%%%%%%%%%%%%%%%%%%%%%%%%%%%%%%%%%%%%%%%
\section*{Acknowledgments}
We thank Concha Gonz\'alez-Garc\'\i a for useful discussions and Steve Blanchet for comments on the manuscript. 

This work has been partially supported by the Spanish MINECO under
grants FPA-2007-60323, FPA-2010-20807,  FPA2011-29678-C02-01, 
Consolider-Ingenio PAU (CSD2007-00060) and CUP (CSD2008-00037), 
by CUR Generalitat de Catalunya grant 2009SGR502
 and by Generalitat Valenciana grant PROMETEO/2009/116.
The work of J.~R.~is also supported by the MINECO Subprogramme Juan de la Cierva. 
M.~P.~is also supported by a FPU-MEC grant. In addition we acknowledge partial support from the  European Union FP7  ITN INVISIBLES (Marie Curie Actions, PITN- GA-2011- 289442).

%%%%%%%%%%%%%%%%%%%%%%%%%%%%%%%%%%%%%%%%%%%%%%%%%%%
\providecommand{\href}[2]{#2}\begingroup\raggedright\endgroup
%%%%%%%%%%%%%%%%%%%%%%%%%%%%%%%%%%%%%%%%%%%%%%%%%%%%%%%%%%%%%%%%%%%%%%%%%%%%%%

\end{document}